\documentclass[11pt]{article}
\newcommand{\be}{\begin{eqnarray}}\newcommand{\beq}{\begin{equation}}
\newcommand{\ee}{\end{eqnarray}}\newcommand{\eeq}{\end{equation}}
\newcommand{\De}{\Delta}

\input epsf
\input amssym.def
\input amssym.tex
\usepackage{graphicx}
\usepackage{times}
\usepackage{helvet}
\usepackage{courier}
\usepackage{cite}
\title{Histogram analysis as a method for determining the line tension 
by Monte-Carlo simulations}
\author{{Y.S.Djikaev}\thanks{E-mail:id45@cornell.edu}\\ 
Department of Chemistry, Baker Laboratory, Cornell University
\\ Ithaca, New York  14853-1301}
\date{}
\begin{document}
\maketitle
\renewcommand{\baselinestretch}{2}\small\normalsize
\begin{abstract}

A method is proposed for determining the line tension, which is  the main physical characteristic of a
three-phase contact region, by  Monte-Carlo (MC) simulations.   The key idea of the proposed method  is that 
if a three-phase equilibrium involves a three-phase contact region, the probability distribution of states of
a system as a function of two order parameters depends not only on  the surface tension, but also on the line
tension. This probability distribution can be obtained as a normalized histogram by appropriate MC
simulations, so one can use the combination  of histogram analysis and finite-size scaling to study the
properties of a three phase contact region.   Every histogram and results extracted therefrom will  depend on
the size of the simulated system. Carrying out MC simulations for a series of system sizes and extrapolating
the results, obtained from the corresponding series of histograms, to  infinite size, one can determine the
line tension  of the three phase contact region and the  interfacial tensions of all three interfaces (and
hence the contact angles) in an infinite system. To illustrate the proposed method, it is applied to the
three-dimensional  ternary fluid mixture, in which molecular pairs of like species do not interact whereas
those of unlike species interact as hard spheres. The simulated results are in agreement with expectations.

\end{abstract}
\renewcommand{\baselinestretch}{2}\small\normalsize
\section{Introduction}

If two bulk phases are in equilibrium, the profiles of the order parameter (distinguishing between the phases)
in the planar interface between them are non-uniform. Due to this non-uniformity, excess contributions to the
extensive thermodynamic parameters of the system arise. These are usually referred to the dividing surface
chosen within the interfacial region. An excess quantity per unit area of the dividing surface is called a
surface adsorption. All surface adsorptions but one depend on the choice of the dividing surface. The special
surface adsorption is that of the grand canonical potential, and is called the surface (or interfacial)
tension; it does not depend on the choice of the dividing surface (in the case of a planar interface). 

\par The interfacial tension between coexisting phases plays a fundamental role in a theory of wetting
phenomena and phase transitions. Various theoretical, numerical, and  computational techniques exist for
its calculation. Besides the mechanical  definition of the surface tension, there are two 
statistical-mechanical recipes  usually referred to as the virial route$^{1}$ and  direct
correlation function route (see, e.g., chapter 4 of ref.2).    The latter is the more rigorous because it
has no restrictions while the former is restricted to fluids with pair-wise additive intermolecular
potentials.  Although not easy in a general case, the equivalence  of both methods can be proven (see,
e.g., chapter 4 in ref.2 and references therein).  These methods require the knowledge
of the pair (and higher order) intermolecular  potentials and the corresponding distribution functions in
the interfacial region, which are not easy to obtain without drastic approximations.$^{2}$ 

\par A more successful and traditional (although semi-numerical) way of calculating  the surface tension
is offered by density functional theory$^{2-4}$ which allows one to find the excess grand canonical
potential associated with the interface between coexisting phases. This method has its
own conceptual difficulties (see, e.g.,refs.3-5) related to  (a) the choice of free-energy density for
intermediate (between two bulk) values of the order parameter, particularly in the critical region where
fluctuations are important, and (b) the long-wavelength interface instability due to capillary
waves.$^{6}$

Finally, there is  a variety of computer simulation techniques for calculating the surface tension. In the
case of a system of continuous potentials the majority of simulations have used the mechanical
definition$^{1}$ of the surface tension and the virial expressions for the normal and tangential components
of the pressure tensor.$^{7-9}$ The results obtained via these simulations show a large scatter.$^{10}$
Another simulation method involves the direct calculation of the free energy change in forming an
interface.$^{11,12}$ In simulation of lattice models a wider variety of methods has been used including
thermodynamic integration using a biasing field to force the formation of the interface$^{13-15}$ and
calculating the free energy differences for systems with periodic and anti-periodic boundary
conditions.$^{16,17}$ 

All these methods implicitly or explicitly include conditions restricting the interfacial  configuration
space that can be sampled in simulations. Many of the methods also require establishing a well-defined
interfacial region which can be quite difficult, particularly under conditions close to a critical point
where the interface becomes very diffuse. An alternative approach to calculating the surface tension was
proposed by Binder,$^{18}$ whose method is based on the analysis of the probability distribution of states
of a system as a function of the order parameter obtained from Monte-Carlo simulations as a normalized
histogram. Under conditions where two phases can coexist in equilibrium, the histogram has two peaks each of
which corresponds to a  homophase state, i.e., to   the whole system being in one of the two phases (Figure
1). Intermediate values of the order parameter correspond to heterophase states such that one  part of the
system is in one phase while the rest is in the other phase. The key idea of  Binder's method consists of
establishing a relation between the minimum of the histogram and the  interfacial tension between two
coexisting phases. A simulated histogram gives the interfacial tension of the finite-size (simulated)
system, and the interfacial tension of the infinite system is obtained by extrapolating a series of results
for systems of different sizes  (but otherwise for the same thermodynamic conditions)  to infinite size.
Binder's method has been successfully applied to a wide variety of models.$^{18-31}$ 

If three phases are in equilibrium, there may exist a three-phase contact region in addition to three
two-phase interfaces. The distortion of the order parameter(s) profile(s) in the three-phase contact region
gives rise to excess contributions to the extensive thermodynamic characteristics of the system.  These are
usually referred to the three-phase contact line". Excess quantities per unit length of the contact
line are called linear adsorptions.  For three bulk phases in equilibrium, the interfaces between them can be
treated as planar and the line of  their intersection, that is, the contact line, as a straight line, so
that its spatial location is determined by two coordinates in  the plane perpendicular to it.  The location
of the contact line is arbitrary and  among all linear adsorptions only one does not depend on its choice.
This is the linear adsorption of the grand canonical potential also  called the line tension.$^{2}$ All 
other adsorptions depend on the location of the contact line and hence are functions of its two coordinates.
It should be noted that line tension may be positive or negative,$^{2}$ unlike  surface tension that is
always positive.

\par The line tension has been studied to a much lesser extent than the surface tension both experimentally
and theoretically.$^{32-34,2,35-40}$    There are no general  statistical mechanical expressions for
calculating the line tension that would be equivalent to the virial  and  direct correlation function
routes in a theory of surface tension.  For the  case of two fluid phases in contact with a solid surface,
a  statistical mechanical theory for the line  tension was proposed by Tarazona and Navascu\'es;$^{41}$
using several assumptions about molecular  interactions, they derived an expression for the line tension
involving the one- and two-molecule distribution functions. All other microscopic models for  the line
tension were  based on the local density approximation of density functional theory$^{2,36-40}$ and are
hence subject to conceptual problems similar to those for the surface tension.$^{3-5}$  As for computer
simulation methods, they have often been applied to three-phase equilibria and wetting phenomena,$^{42-46}$
but we are not aware of direct molecular simulations  of the line tension, which is probably  partly due to
the lack of a general  statistical-mechanical theory thereof. It should be noted, however, that  studying a
solid spherical particle at a liquid-vapor interface by molecular dynamics (MD) simulations and thus
finding the contact angle and all interfacial tensions involved, Bresme and Quirke$^{43}$ evaluated the
(particle-liquid-vapor) line tension with the help of a phenomenological expression for the particle free
energy (found also by MD simulations) and modified Young's equation.

\par In the present paper a method is proposed for calculating the line tension by MC simulations. The
method  
is based on the statistical-mechanical  analysis of the histogram of a system under conditions of three phase
equilibrium involving a three-phase contact region.  The paper is structured as follows. Section 2 outlines
the original histogram analysis method for determining the interfacial tension of a two-phase interface
pioneered by Binder.$^{18}$  Section 3  presents a general idea of how that method can be extended to
three-phase equilibria in order to determine the line tension. Section 4 discusses a general set-up of MC
simulation cell and boundary conditions and there is also developed a theoretical basis for the  concrete
realization of the key  idea in a three-dimensional 
ternary fluid mixture, in which molecular pairs of like species do not interact while those of unlike
species interact as hard spheres;$^{47-49}$ in this model a three-phase contact region may exist 
at densities above the ``quadruple point"  density.$^{47-49}$  
The details of MC simulations and  results for
the surface and line tensions  associated with the three-phase contact region  are also presented in section
4.  The results are discussed and  conclusions are summarized in section 5. 

\section{Binder's approach to determining the surface tension by Monte-Carlo simulations} 

Binder's method$^{18}$ for determining the interfacial tension of two coexisting phases   is based on the
order-parameter distribution formalism. Rather than set up an explicit interface in the MC-simulated
system, this approach relies on spontaneous fluctuations that give rise to density inhomogeneities, which
provide information about the interfacial properties.  The method has its own implementation
difficulties,$^{29,31}$  especially far away from the critical point, but they can all be  alleviated by
using special sampling techniques.$^{23,31,50}$ 

\par Consider a finite ``cube-shaped" system (to be simulated) of volume $L^d$ ($d$ is the dimensionality) 
with  periodic boundary conditions (PBC),  which are necessary in order to avoid
additional surface effects that can arise from the system boundaries. Under appropriate thermodynamic
conditions (e.g., below the critical temperature if the temperature is a relevant thermodynamic parameter)
the system will phase separate into phases $\alpha$ and $\beta$ with the values $m_{\alpha}$ and
$m_{\beta}$ of the order parameter $m$.  In a liquid-gas system the order parameter is the density; in a
binary mixture of species $A$ and $B$ with a miscibility gap the order parameter is the mole fraction of
one of the components in the coexisting $A$-rich and $B$-rich phases; in an Ising model, the order
parameter is the spontaneous magnetization (two orientations thereof corresponding to the two phases). In
all cases some conjugate field variable (the chemical potential, or the difference in chemical potentials
of two components, or the magnetic field, respectively) has to be fixed at such a value that  would
correspond to coexisting phases in the thermodynamic limit. 

\par According to the theory of equilibrium fluctuations,$^{51}$ if the system is in a pure phase with
the average order parameter $\bar{m}$, the probability $p(m)$ 
to find the system (with $L\rightarrow \infty$) 
in a state with the order parameter $m$ is given by the Gaussian distribution,
\beq p(m)=L^{d/2}(2\pi kT\chi)^{-1/2}\exp[-(m-\bar{m})^2L^d/(2kT\chi)],\eeq
where $k$ is the Boltzmann constant, $T$ is the temperature, and $ \chi$ is the partial derivative of the
order parameter with respect to its conjugate field variable, with 
all other variables of state of the system 
being fixed. For the system at coexistence where both pure phases are equally likely to occur, 
the probability distribution of the order parameter can be represented as the sum of 
two displaced Gaussians,
\be p(m)&=&\frac1{2}L^{d/2}(2\pi
kT\chi_{\alpha})^{-1/2}\exp[-(m-m_{\alpha})^2L^d/(2kT\chi_{\alpha})]\nonumber\\
&+&\frac1{2}L^{d/2}(2\pi kT\chi_{\beta})^{-1/2}\exp[-(m-m_{\beta})^2L^d/(2kT\chi_{\beta})],\ee
where $\chi_{\alpha}$ and $\chi_{\beta}$ are the susceptibilities in the pure phases $\alpha$ and $\beta$. 
It is clear from eq.(2) that the probability of a {\em homogeneous} state with the order parameter 
$m_{\mbox{\tiny min}}=(m_{\alpha}+m_{\beta})/2$ decreases exponentially with the volume of $L^d$ of the 
system. On the other hand, the probability of {\em heterogeneous} fluctuations, such that 
two phases coexist in the system, decreases exponentially$^{18,56}$ with the interface area $2L^{d-1}$,
\beq p(m^{\mbox{\tiny ht}}_{\mbox{\tiny min}})\propto p(m_{o})\exp[-2\sigma L^{d-1}/kT],\eeq
where the superscript ``ht" stands for ``heterogeneous", the subscript ``{\em o}" indicates  either of the
pure phases $\alpha$ or $\beta$, and $\sigma$ is the interfacial tension  in an infinite system. The
size dependence of the pre-exponential factor  $p(m_{o})\simeq \frac1{2}L^{d/2}(2\pi kT\chi_o)^{-1/2}$  is
much weaker than the exponential one. Thus, for large enough $L$, the probability of  ``homophase"
fluctuations with intermediate values $m_{\mbox{\tiny min}}$ of the order parameter is  negligible in
comparison with the probability of ``heterophase" fluctuations. 

\par As pointed out by Binder,$^{18}$  the size dependence of the pre-exponential factor in eq.(3) can be
more complicated than that contained in  $p(m_{o})$, due to finite-size contributions to the interfacial
free energy arising, e.g., from the ``Goldstone modes" and capillary waves. As a result, one can expect that
for large $L$ the probability distribution  $p(m_{\mbox{\tiny min}})$ (the superscript ``ht" is omitted
henceforth) has the general form

\beq p(m_{\mbox{\tiny min}})=aL^{x}p(m_{o})\exp[-2\sigma L^{d-1}/kT] \eeq
with unknown $a$ and $x$, whence it follows that
\beq \frac1{2L^{d-1}}\ln\frac{p(m_{o})}{p^(m_{\mbox{\tiny min}})}=\frac{\sigma}{kT}+
b\frac{\ln L}{L^{d-1}}+c\frac1{L^{d-1}}, \eeq 
(with unknown $b$ and $c$) and  
\beq \frac{\sigma}{kT}=\lim_{L\rightarrow\infty}
\frac1{2L^{d-1}}\ln\frac{p(m_{o})}{p(m_{\mbox{\tiny min}})}.\eeq
This limit cannot be found analytically, but   the probability distribution as a function of the order
parameter, $p(m_{o})$,  can be obtained for a series of $L$'s via appropriate MC simulations.
Then, one can plot the LHS of eq.(5) vs  $\ln L/L^{d-1}$ and the intercept of this plot with the ordinate
axis will provide the ratio  $\sigma/kT$ for the interfacial tension in the infinite system. 

\section{Three-phase equilibrium: determination of the surface and line tensions by the histogram analysis 
method}  
If the system is under such conditions that three phases can coexist in equilibrium, there are 
necessarily two
order parameters such that at least one in the pair differs in different phases. For example, in a single
component system at the triple point the gas, liquid, and solid phases can coexist, and a pair of order
parameters can be the density and a  structure parameter. In some two component systems 
(``solvent-solute") under appropriate conditions  one can observe the coexistence of  three fluid phases:
solvent liquid, solvent vapor, and solute liquid.$^{30}$  In such systems the densities of the  components
serve as a pair of order parameters. For such systems, the probability  distribution of states is a function of
two independent order parameters and determines a three-dimensional surface, hereafter referred to as a
histogram (as in a single order parameter case).  This three-dimensional surface has three 
peaks of equal height corresponding to three pure phases.

\par Consider a  three component fluid system of species $1$, $2$, and $3$. Under appropriate conditions, such
a system can  phase-separate (demix) into three  equilibrium phases differing from each other by a pair of
independent  order parameters, e.g., the mole fractions of any two components, say, $m_1$ and $m_2$  (the
third mole fraction is not an independent variable, $m_3=1-m_1-m_2$).  Denote by $p(m_1,m_2)$ the probability
distribution of states of this system with respect to the order parameters $m_1$ and $m_2$.  Figure 2a
presents an example of the histogram (three dimensional surface determined by the function $p(m_1,m_2)$) 
generated by semigrand canonical MC simulations in a cubic box with PBC  for the  three-dimensional  ternary
fluid mixture, in which molecular pairs of like species do not interact and those of unlike species interact
as hard spheres (for details see section 4).  Figure 2b shows the projection of the same surface onto the
mole-fraction plane.  The peaks of the histogram correspond to the pure phases, say, $\alpha$, $\beta$, and
$\gamma$.  The projections of the peaks are shown as the points $o_1;\;\;(o_1=\alpha,\beta,\gamma)$ in
Fig.2b.   Their coordinates $m^{o_1}_i\;\;(i=1,2)$ determine the composition (i.e. mole fractions)  of the
system in pure phases $\alpha$, $\beta$, and $\gamma$. If all the three phases are symmetric (as, e.g., for
the model represented by the histogram in Fig.2),
$$m_1^{\alpha}=m_2^{\beta}=m_3^{\gamma},\;\;\; 
m_2^{\alpha}=m_3^{\alpha}=m_1^{\beta}=m_3^{\beta}=m_1^{\gamma}=m_2^{\gamma},$$ 
but this may not be the case in general. 

\par Suppose the system is in some pure phase $o_1\;\;\;(o_1=\alpha,\beta,\gamma)$.  According to the
thermodynamic theory of equilibrium fluctuations,$^{51}$  the probability distribution
of homogeneous fluctuations of order parameters $m_1,m_2$ is 
\be p(m_1,m_2)&=&\sqrt{\det(\kappa)}\frac{V}{2\pi kT/\rho}\exp\left[-
\frac{V}{2kT/\rho}(\kappa_{11}(m_1-m_1^o)^2\right.\nonumber\\
&+&\left. 2\kappa_{12}(m_1-m_1^o)(m_2-m_2^o)+\kappa_{22}(m_2-m_2^o)^2)\right],\ee
where $V=L^d$ is the volume of the system. In a semigrand canonical ensemble (SGCE), which we will
use for our MC simulations,  $\kappa$ is a $2\times 2$ diagonal matrix with elements
\beq \kappa_{ij}=\left(\frac{\partial m_i}{\partial \Delta\mu_j}\right)_{NVT\Delta\mu'}\eeq
with $\De\mu_i=\mu_i-\mu_3\;\;(i=1,2)$ and $\mu_i$ the chemical potential of component $i$.  The prime in
eq.(8) indicates  that one of the $\De\mu$'s is kept constant in the partial derivative. The choice of 
SGCE is the matter of convenience and  efficiency of MC simulations of  phase
separation in dense fluids$^{52-54}$ (see subsection 4.2 for more details). Note that, in the 
thermodynamic limit, SGCE simulation at fixed $V,T,N,\De\mu_1,\De\mu_2$ is
equivalent to grand canonical simulation at fixed $V,T$, and such $\mu_1,\mu_2,\mu_3$ (fixed) that
$\bar{N}$, the grand-canonical  average number of molecules in the system, is equal to $N$.

\par As clear from eq.(7), the probability of a homogeneous state with one or both order parameters having
their intermediate values,   $m_{1\mbox{\tiny min}}=(m_1^{\alpha}+m_1^{\gamma})/2$ and $m_{2\mbox{\tiny
min}}=(m_2^{\beta}+m_2^{\gamma})/2$,  decreases exponentially with the system volume. Hence, for large
enough volumes, ``homophase" fluctuations with $m_1=m_{1\mbox{\tiny min}}$ or/and $m_2=m_{2\mbox{\tiny
min}}$ are negligible compared with ``heterophase" fluctuations. To show this, we will consider the system 
(to be simulated) that has the form of a cylinder of length $L$ and radius $R$ (Figure 3). Such a shape
will be most convenient (but not necessary) to determine the line tension by computer simulations, but
otherwise it does not affect measurable physical quantities because they are  obtained by finite-size
scaling,$^{18,19,31,53-55}$ which involves their extrapolation to infinite cell sizes.

\subsection{Determination of the surface tension}
\par For ``heterophase" fluctuations, intermediate points on the lines ($\alpha\beta$), ($\beta\gamma$),
and  ($\gamma\alpha$)  in Figure 2b  represent the two-phase states of the system with none of the third
equilibrium phase present. For example,  for a point on the line ($\alpha\beta$) phases $\alpha$ and
$\beta$ coexist while the amount of phase $\gamma$ is 0, although thermodynamically it can coexist with the
$\alpha$ and $\beta$. The middle points  $o_{\alpha\beta}=\{m_{1\mbox{\tiny min}},m_{2\mbox{\tiny
min}}\}$,  $o_{\beta\gamma}=\{m_{1}^{\gamma},m_{2\mbox{\tiny min}}\}$, and 
$o_{\alpha\gamma}=\{m_{1\mbox{\tiny min}},m_{2}^{\gamma}\}$ on these three lines correspond to the
situations when one half of the system is in one phase and the other half in another phase. For example,
the point $o_{\alpha\beta}$   corresponds to such s situation that one half of the system is in phase
$\alpha$ and the other half in phase $\beta$, while the equilibrium phase $\gamma$ is not present at all. 
Let us use the generic symbol $o_2$ to denote the points
$o_{\alpha\beta},o_{\beta\gamma},o_{\gamma\alpha}$. The difference between  the free energies of the system
in states $o_1$ and $o_2$ is due only to an interfacial contribution $F_s$ from the interface between two
coexisting phases in the state $o_2$. According to the principles of equilibrium thermodynamics,  $p(o_2)$
and $p(o_1)$, the probability densities of two-phase states $o_2$ and pure phase states $o_1$,  are
related as   
\beq p(o_2)=p(o_1)e^{- F_s/kT}.\eeq   
(phase $o_1$ is one of two phases in the state $o_2$)

\par  There is no exact analog of PBC for 
simulations in a cylindrical geometry, but there are other boundary conditions attenuating the effect of
cell boundaries. This issue will be discussed in section 4, but for now it need only be noted 
that one can impose  PBC along  the cylinder axis (hereafter called the $z$-axis) and assume that the 
effects of the lateral cell boundaries can be neglected.  Because of PBC along the $z$-axis, there may be 
only three principally different  ``half-and-half" spatial  distributions of two equilibrium phases in the
cylinder (Figure 3). Comparing the interfacial free energy contributions $F_s$ in these three cases, one
can show that if the aspect ratio $\omega\equiv L/R$ of the cylinder satisfies the inequalities
\beq   1<\omega<\pi,   \eeq
then the phase partitioning shown in Fig.3a is thermodynamically most favorable. 
Hereafter,  this will be referred to as the axial ``longitudinal" two-phase (AL2P) partitioning. The
aspect ratio $\omega<1$ favors the ``droplet-like" phase partitioning (Fig.3b), while 
$\omega>\pi$ favors the ``transversal" phase  partitioning (Fig.3c). For the droplet-like phase
partitioning the interfacial tension is assumed to be the same as that of a
planar interface. This should be a reasonable assumption for  a large enough droplet.

\par If the histogram is obtained by simulations in a cylinder subject to conditions (10) and PBC along the
$z$-axis, one can represent eq.(9) in the form analogous to eq.(4),
\beq p(o_2)=aL^xR^yp(o_1)e^{-( 2RL \sigma_{o_2} )/kT}.\eeq
Here, $\sigma_{o_2}$ is the surface tension of a flat interface in an infinite system, and the
pre-exponential factor $aL^xR^y$ (with unknown $a,x,y$) is due to finite size effects (Goldstone modes,
capillary waves, etc$^{18}$) that may contribute to $F_s$,  where the leading term 
$2RL\sigma_{o_2}$ is larger than finite size corrections by an order of magnitude. From eq.(11) it follows
that 
\beq \frac1{2RL}\ln\frac{p(o_1)}{p(o_2)}=\frac{\sigma_{o_2}}{kT} 
-x\frac{\ln L}{2RL}-y\frac{\ln R}{2RL}-\frac{\ln a}{2RL}. \eeq
Again, for an infinitely large system the RHS of this equation will equal the interfacial tension. The
limit $L\rightarrow\infty,\;R\rightarrow\infty$ can be reached in such a way that  $\omega(=L/R)$
remains constant, so that equation (12) can be rewritten in two forms, 
\be \frac1{2RL}\ln\frac{p(o_1)}{p(o_2)}&=&\frac{\sigma_{o_2}}{kT} 
+b\frac{\ln L}{L^2}+c\frac{1}{L^2}\nonumber\\
&=& \frac{\sigma^{o_2}}{kT} + d\frac{\ln R}{R^2}+f\frac{1}{R^2}\ee
with unknown $b,c,d,f$. The probability distribution as a function of the order parameters, $p(m_1,m_2)$, 
can be  obtained for a series of $L$'s (or $R$'s) via appropriate MC simulations.  Then, one can
plot the LHS of eq.(13) vs $\ln L/L^2$ (or vs $\ln R/R^2$) and  the ordinate intercept of this plot will
provide the ratio  $\sigma_{o_2}/kT$ for the interfacial tension in an infinite system. 

\subsection{Determination of the line tension}

\par ``Heterophase" states, corresponding to points lying off the two-phase lines ($\alpha\beta$),
($\beta\gamma$), and  ($\gamma\alpha$)  in Figure 2b, represent three-phase states such that all three
equilibrium phases are simultaneously present in the system. The amount of each phase present is determined
by the coordinates of a point $o'_3$ corresponding to a  given  three-phase state.  When the three
equilibrium phases are symmetric,  the central  point $e_3$  of the
composition triangle corresponds to  a  state in which all three phases are present  in equal volumes. 
In a general case this three-phase volume equi-partition point is not necessarily in the middle of
the composition triangle. 
\par Since the subject of the present work  is the line tension, we restrict our
interest to three-phase equilibria involving three-phase contact regions. The necessary and sufficient
condition that there exists such a region in a three-phase system is 
\beq \sigma_{\alpha\gamma}<\sigma_{\alpha\beta}+\sigma_{\beta\gamma},\eeq  
(hereafter double Greek subscripts and superscripts indicate quantities for corresponding two-phase
interfaces). Condition (14) means that it is  thermodynamically unfavorable for the system to have three
phases arranged in three  layers parallel to each other.  When this condition is fulfilled, $F_{sL}$, the
difference between  the free energies of the system in states $o_1$ and $o'_3$ may contain not only
interfacial contributions  from three two-phase interfaces, but also the ``line tension" 
contribution from the three-phase contact region. Thus, $p(o'_3)$ and $p(o_1)$, the probability
densities of three-phase states $o'_3$ and pure phase states $o_1$  are related as 
\beq p(o'_3)=p(o_1)e^{-F_{sL}/kT}, \eeq
where the excess free energy $F_{sL}$ has the form 
\beq F_{sL}=\sigma_{\alpha\beta}A^{\alpha\beta}+\sigma_{\beta\gamma}A^{\beta\gamma}+
\sigma_{\alpha\gamma}A^{\alpha\gamma}+\tau L^{\alpha\beta\gamma}.\eeq
Here the $A$'s are the surface areas of interfaces and  
$L^{\alpha\beta\gamma}$ is the length of the three phase contact line (assumed to be straight) with 
which the line tension $\tau$ is associated. 

\par Again consider a cylindrical simulation cell with the boundary conditions discussed above, and assume
that there exists $1<\omega_0<\pi$ such that if $L/R=\omega_0$, then
thermodynamically the most favorable three-phase partitioning is the ``longitudinal" one shown 
in Figure 4a. When the three-phase contact line coincides with the axis of the cylinder (as shown
in  Fig.4a), this will be referred to as the axial longitudinal three-phase (AL3P) partitioning. 
Denoting by $o_3$ the corresponding point in the plane $m_1,m_2$, one can rewrite eq.(15) in the form 
\beq 
p(o_3)=aL^xR^yp(o_1)e^{-[(\sigma_{\alpha\beta}+\sigma_{\beta\gamma}+\sigma_{\alpha\gamma})LR+\tau L]/kT},
\eeq 
where now $\sigma$'s and $\tau$ are the interfacial and line tensions, respectively, in an infinite systems, 
and the pre-exponential factor $aL^xR^y$ (with unknown $a,x,y$, in general different from those in
eq.(11)) is again due to finite size effects: because $\sigma$'s and $\tau$ in eqs.(16) and 
(17) are those of an infinitely large system, there may arise small (in comparison with the leading terms) 
finite-size corrections to $F_{sL}$. 

\par Henceforth assuming that the simulations are carried out under such conditions that the
AL2P and AL3P partitionings are thermodynamically most favorable, 
one can combine eqs.(11) and (17) and obtain
\beq p(o_3)=aL^xR^y[p(o_1)]^{1-\xi}[p(o_2)]^{\xi}e^{-\tau L/kT}, \eeq 
where $a,x,y$ are unknown (different from those in eqs.(11) and (17)) and 
\beq \xi=\frac{\sigma_{\alpha\beta}+\sigma_{\beta\gamma}+\sigma_{\alpha\gamma}}{2\sigma_{o_2}}.\eeq
From eq.(18) it follows that 
\beq \frac1{L}\ln\frac{[p(o_2)]^{\xi}}{[p(o_1)]^{\xi-1}p(o_3)}=\frac{\tau}{kT} 
-x\frac{\ln L}{L}-y\frac{\ln R}{L}-\frac{\ln a}{L}.\eeq
For an infinitely large system the RHS of eq.(20) will equal the ratio of the line tension $\tau$ 
to $kT$. If the limit $L\rightarrow\infty,\;R\rightarrow\infty$ is reached in such a way that 
$\omega\equiv L/R$ remains constant, then eq.(20) can be rewritten as 
\be \frac1{L}\ln\frac{[p(o_2)]^{\xi}}{[p(o_1)]^{\xi-1}p(o_3)}&=&\frac{\tau}{kT} 
+b\frac{\ln L}{L}+c\frac{1}{L}\nonumber\\
&=& \frac{\tau}{kT} 
+d\frac{\ln R}{R}+f\frac{1}{R}\ee
with unknown $b,c,d,f$. Again, $p(m_1,m_2)$, the probability distribution as a function of the order
parameters, can be  obtained for a series of $L$'s (or $R$'s) via appropriate Monte-Carlo simulations at
constant  $\omega(=L/R)$.  Then, one can plot the LHS of eq.(19) vs $\ln L/L$ (or vs $\ln R/R$) and  the
ordinate intercept of this plot will provide the ratio $\tau/kT$ for  
the line tension in the infinite system. 
\par It is worth emphasizing that the 
point $o_1$ in eqs.(9)-(21)  corresponds to any pure phase state (${\alpha},{\beta}$,
or ${\gamma}$) because (ideally) in a three-phase equilibrium 
$p(m_1^{\alpha},m_2^{\alpha})=p(m_1^{\beta},m_2^{\beta})=p(m_1^{\gamma},m_2^{\gamma})$.  The point $o_2$
represents any two-phase equilibrium with the AL2P partitioning (Fig.3a) involving $o_1$ phase, and 
$\sigma_{o_2}$ in eqs.(11)-(13),(19) is the corresponding interfacial tension. Finally, the point $o_3$
in eqs.(17)-(21) represents the AL3P partitioning (Fig.4a). 

\section{General set-up of the simulation cell and boundary conditions} 
\par For simulations in a rectangular geometry one usually uses PBC in order to decrease the effect from the
simulation cell boundaries. For  simulations in a cylindrical geometry, there is no exact analog of PBS, but 
one can impose boundary conditions that do attenuate the effect of cell boundaries. Namely, one can  impose
the usual PBC along the  $z$-axis (Figures 3 and 4). In the plane of polar coordinates $\phi,r$, perpendicular
to the $z$-axis, impose  ``random-periodic" boundary conditions (RPBC); that is, if upon an elementary
Monte-Carlo move the  {\em new} coordinates  $z,\phi,r$ of a molecule are such that $r>R$ (i.e., the molecule
crosses through the lateral surface of the cylinder), the angle $\phi$ is replaced by a new, {\em random} angle
$\phi'$ and the radius $r$ is replaced by  a new, {\em periodic} value $r'=R-(r-R)$. After these changes are
made, the acceptance criterion is checked and the move is either accepted or rejected. If the move is rejected,
the molecule keeps its old coordinates (those prior to generating new  coordinates $z,\phi,r$). In this work
only hard-sphere type  molecular interactions are considered,  but in a general case (e.g., Lennard-Jones
intermolecular potential) it would  be necessary to deal with the problem of accounting for the interactions
of  ``random-periodic" images of molecules, that are in the lateral surface layer of ``cutoff" thickness, with
other molecules in this layer.  

\par The propensity of an equilibrium system to minimize its free energy and  PBC along the $z$-axis
prevent the formation of artificial (i.e., due to the finite size of the simulation cell) phase interfaces
at the bases of the cylinder. However,  imposing RPBC in the $\phi,r$ plane gives rise to the
{\em probabilistic}  phase interfaces at the lateral surface of the cylinder. Actually, with RPBC imposed,
a molecule crossing the lateral
interface will leave phase $s_1$ and enter phase $s_2$ with the probability  $g_{s_1s_2}$ 
($s_1,s_2=\alpha,\beta,\gamma$), 
\beq g_{s_1 s_2}=g_{s_2 s_1}=\frac{s_1 s_2}{(2\pi)^2}\;\;\;\;\;(s_1,s_2=\alpha,\beta,\gamma), \eeq 
(where $s_1$ and $s_2$ are not subscripts, they must be understood as the corresponding contact angles) 
with the normalization condition $\sum_{} g_{s_1 s_2}=1$ naturally satisfied.  
Thus, the whole lateral surface of  area $2\pi RL$ is an $\alpha\beta$ interface with the probability
$2g_{\alpha\beta}$, a $\beta\gamma$ interface with the probability $2g_{\beta\gamma}$, and an $\alpha\gamma$
interface with the probability $2g_{\alpha\gamma}$.
Clearly, for any point $o_2$ (half-and-half two-phase coexistence) the analogous probability is 
$2g_{o_2}=1/2$. 

\par Thus, as an artifact of RPBC imposed at the lateral surface of the simulation
cell (cylinder), it is necessary to modify equations (11)-(13), (17), and (19). Namely, eqs.(11)-(13) must 
be replaced by 
\beq p(o_2)=aL^xR^yp(o_1)e^{-  (2+\pi)RL \sigma_{o_2}/kT},\eeq 
\beq \frac1{(2+\pi)RL}\ln\frac{p(o_1)}{p(o_2)}=\frac{\sigma_{o_2}}{kT}
-x\frac{\ln L}{(2+\pi)RL}-y\frac{\ln R}{(2+\pi)RL}-\frac{\ln a}{(2+\pi)RL}, \eeq 
\be \frac1{(2+\pi)RL}\ln\frac{p(o_1)}{p(o_2)}&=&\frac{\sigma^{o_2}}{kT}
+b\frac{\ln  L}{L^2}+c\frac{1}{L^2}\nonumber\\ 
&=& \frac{\sigma_{o_2}}{kT}+d\frac{\ln R}{R^2}+f\frac{\ln R}{R^2},\ee
respectively. Equations (17) and (19) must be replaced by  
\beq p(o_3)=aL^xR^yp(o_1)e^{-[((1+4\pi
g_{\alpha\beta})\sigma^{\alpha\beta}+ (1+4\pi g_{\beta\gamma})\sigma_{\beta\gamma}+(1+4\pi
g_{\alpha\gamma})\sigma^{\alpha\gamma})LR +\tau L]/kT}, \eeq 
\beq \xi=\frac{(1+4\pi 
g_{\alpha\beta})\sigma_{\alpha\beta}+ (1+4\pi g_{\beta\gamma})\sigma_{\beta\gamma}+(1+4\pi
g_{\alpha\gamma})\sigma_{\alpha\gamma}} {(2+\pi)\sigma_{o_2}}.\eeq 
Now, the  ratio $\sigma_{o_2}/kT$ for the surface tension of an infinite system is determined as the
ordinate intercept of the plot of the LHS side of eq.(25) vs  $\ln L/L^2$ (or vs $\ln R/R^2$).  The
ratio $\tau/kT$ for the line tension in an infinite system is found, as previously, by plotting the 
LHS of eq.(21) vs $\ln L/L$ (or vs $\ln R/R$) and finding  the ordinate intercept of this plot, 
but the parameter $\xi$ is now given by eq.(27)  rather than eq.(19). 

\par Note that even with perfect PBC imposed on the finite simulation cell, there may  be cell {\em boundary}
contributions to $F_s$ and $F_{sL}$, but they can be expected to be much smaller than the above contributions
from the {\em physical} phase interfaces at the cell boundaries.  Moreover, in the case where the equilibrium
phases are symmetric (such as Ising-like models, the model discussed below, etc), the cell {\em  boundary}
contributions to $F_s$ (and $F_{sL}$) can be assumed to be zero$^{18,55}$ because in such systems the free
energy of the cell is affected by its boundaries in the same way, independently of which phases are present. 

\subsection{Method specifics for the simulated model}

To illustrate the proposed method with a concrete example, MC simulations were carried out for  the
three-dimensional three-component fluid mixture, in which   the molecules of different species interact as
hard spheres, whereas molecules of the same species do not interact at all (i.e., are completely penetrable by
one another).$^{47-49}$   This is a natural generalization of the binary model, earlier proposed by Widom and
Rowlinson$^{56}$  and come to be known as  the primitive version of the Widom-Rowlinson model. 
The intermolecular potentials between two molecules of species $i$ and $j$ at a distance $r$ from each other 
are defined as 
\beq u_{ii}(r)=0\;\;\;(i=1,2,3),\eeq 
\beq u_{ij}(r)=0\;\;\;(r>\eta,i\ne j=1,2,3),\;\;\;\;\;\; u_{ij}(r)=\infty\;\;\;(r<\eta,i\ne j=1,2,3),\eeq 
so that the molecules of each pure species constitute an ideal gas, while every molecule of species $i$  sees
every molecule of species $j\;\;\;(j\ne i)$ as a hard sphere of diameter $\eta$. The phase behavior of this
system was analytically studied in both mean-field and higher order approximations.$^{47-49}$ The  system is
predicted to have a so-called ``quadruple point" density $\rho_q$: at this density within  the composition
triangle there is a smaller one (similar to and co-centered with the former)  where an equimolar ternary
mixture may exist in  equilibrium with three symmetric phases each rich in one of three components.   When the
density becomes  greater than $\rho^q$, the equimolar phase is no longer present and the inner four-phase
triangle becomes a  three-phase coexistence region. The three coexisting phases $\alpha$, $\beta$, and
$\gamma$ are related symmetrically: phase $\alpha$ is rich in component $1$  with equal traces of $2$ and
$3$,  phase $\beta$ is rich in $2$ with equal traces of $1$ and $3$, and  phase $\gamma$ is rich in $3$ with
equal traces of $1$ and $2$.

\par Since the three coexisting phases $\alpha,\beta$, and $\gamma$ are absolutely symmetric, the three
interfacial tensions are all equal, 
$\sigma^{\alpha\beta}=\sigma^{\beta\gamma}=\sigma^{\alpha\gamma}=\sigma$.  Therefore a three-phase coexistence
will necessarily involve a three phase contact region with the contact angles $\alpha=\beta=\gamma=2\pi/3$.
Furthermore, for the appropriate size and aspect ratio $\omega_0$  of the cylindrical simulation cell, the
point $o_3$ which corresponds to the AL3P  partitioning will coincide with the phase volume equi-partition
point $e_3$, and both $o_3$ and $e_3$ lie exactly in the center of the three-phase triangle. 

\par As  an alternative to AL3P, the three-phase partitioning may be of
mixed character as illustrated in Fig.4b (for $\omega>\omega'=(6-8/\pi)/\sqrt{3}$) and  Fig.4 
(for $\omega<\omega'$). Knowing the contact angles ($\alpha=\beta=\gamma=2\pi/3$) enables one to determine
the conditions under which in  the state $e_3$ (=$o_3$) the AL3P partitioning (Fig.4a) 
is thermodynamically most favorable:  
\beq \frac{s_{}(\omega)}{4-\omega}>-\frac{\tau}{\sigma R}\;\;\;\;(1<\omega\le\pi),\eeq 
(only the range $1<\omega<\pi$ is considered because of condition (10) for the AL2P partitioning).
The explicit form of the continuous function $s(\omega)$ is given in the Appendix.  The function
$s_{}(\omega)/(4-\omega)$ is positive for $1<\omega<\omega^*\simeq 2.465$, attaining  its maximum at $\omega
\simeq 1.635$. Thus, if $\tau>0$ and $1<\omega\lesssim \omega^*$, the AL3P partitioning (shown in Fig.4a) is
thermodynamically more favorable than those in Fig.4b and Fig.4c.  When  $\tau<0$ (as the mean-field theory
predicts$^{36}$ for the 3d3c-PS model) and $1<\omega\lesssim \omega^*$, making the AL3P partitioning 
thermodynamically most favorable requires that $R$ be sufficiently large, 
$$R>R^*=\frac{4-\omega}{s(\omega)}\frac{|\tau|}{\sigma}.$$ 

\par Evaluating $R^*$ accurately prior to simulations is complicated by the absence of  data on $\tau$, even
though data on $\sigma$  were readily available.  Besides, taking large $R$'s (for a given  density) leads 
to a significant increase in  the CPU time for MC simulations.  If this is not an acceptable option,
one can use a modified  version of the above method. For the case of a mixed three-phase partitioning,  the
modifications  lead to the equations  
\beq p(o_3)=aR^x[p(o_1)]^{1-\xi}[p(o_2)]^{\xi}e^{-4R\tau /kT}, \eeq   
\beq \xi=\frac{\phi(\omega)}{(2+\pi)\omega},\eeq 
\be 
\frac1{4R}\ln\frac{[p(o_2)]^{\xi}}{[p(o_1)]^{\xi-1}p(o_3)}&=&
\frac{\tau}{kT}+b\frac{\ln R}{R}+c\frac{1}{R} \nonumber\\
&&\frac{\tau}{kT}+d\frac{\ln L}{L}+f\frac{1}{L} \ee
where $a,x,b,c,d,f$ are unknown and the explicit form of the function $\phi(\omega)$ is given
in the Appendix.  Generating $p(m_1,m_2)$, the probability distribution as a function of the order
parameters,  via appropriate MC simulations for a series of $R$'s (or $L$'s) at constant $\omega=L/R$, 
and plotting the LHS of eq.(33) vs $\ln R/R$ (or $\ln L/L$), one finds  the ratio $\tau/kT$ 
as the ordinate intercept of this plot. 

\subsection{Monte-Carlo procedure and results}

To obtain the histogram of a system showing a demixing phase transition, most convenient are  SGCE MC
simulations$^{57-59}$, where  it is necessary to explore all configurations and compositions of the system for
a constant total number of molecules. The configurational space is sampled as in a standard canonical
simulation. Varying the composition is carried out by the so-called ``identity change" MC moves, and can be
done in two different ways: (1) choose a molecule at random, change its identity, and accept or reject the
change; (2) choose a component at random, choose a molecule of that component at random, change its identity,
and accept or reject the change. These two methods of composition sampling correspond to two different ways to
express the SGCE partition function.$^{57}$ As a consequence, the acceptance criteria for these methods are
different. As a slight modification of this algorithm, an identity change attempt may be combined with the
configurational displacement of a molecule.$^{60,61}$ The choice between a pure spatial MC displacement and a
combined ``spatial displacement+identity change" MC move is uniformly random.  

\par The most obvious advantage of SGCE MC simulations over the Gibbs ensemble simulations$^{63-65}$ is 
that less computational effort is required by the former (single phase simulations of $N$ particles)
compared to the latter (two-phase simulation of $N+N$ molecules). Besides, the Gibbs ensemble method
relies on particle transfers from one fluid phase (=simulation box) into another, which gives rise to
technical limitations when studying relatively dense phases, because the probability of particle insertion
is then very low and the method becomes inefficient. Similar and other problems arise in a grand
canonical ensemble  simulations of dense fluids, although recently some special techniques have been
developed to overcome or alleviate those problems$^{23,30,31,51,55}$ 

\par As a test of the simulation algorithm, the SGCE MC simulations were carried out for the primitive version
of the two-dimensional 
Widom-Rowlinson model$^{56}$ in a square cell with standard PBC and the critical density was determined 
by using the  method proposed by Binder and co-workers.$^{19,52,53,62}$ They showed that for demixing phase
transitions in two-component mixtures, the ratio $<s^2>/<|s|>^2$ (the reduced second moment  of $s=2m-1$, with 
$m$ being the order parameter) at the critical density $\rho_c$ 
is independent of the system size. Hence, $\rho_c$ in
such systems can be determined as the density at which the curves ``$<s^2>/<|s|>^2$ vs density" for
different system sizes cross.  The simulations were carried out   for four different total  numbers of
molecules in the system: $N=256,\;726,\;1024,$ and $4098$. The results are shown in Figure 5, where
$<s^2>/<|s|>^2$ is plotted as a function of the dimensionless density $\rho\eta^2$ for four $N$'s. The
critical density was determined  as the average of six densities at which six pairs of simulated series
cross.  The standard deviation of this scatter dominates the standard deviation of $<s^2>/<|s|>^2$ for any
given density in a single simulation run. Thus for a two-dimensional Widom-Rowlinson model the dimensionless 
critical 
density was found to be  $\rho_c\eta^2\simeq 1.58 \pm 0.01$. This is very close to the  result reported
previously by Johnson et al.,$^{66}$ who used the invaded cluster  (IC) approach in MC simulations: 
$\rho_c\eta^2\simeq 1.566\pm 0.003$. To our knowledge, there have been no other simulations of this model so
far.

\par The MC simulations of the three-component three-dimensional model 
were carried out in a cylindrical  cell with PBC along the $z$-axis
and RPBC at the lateral surface of the cell. Two  densities, $\rho\eta^3=0.83$ and $\rho\eta^3=0.85$, were
considered, both  slightly higher than the quadruple point density of this model reported
previously$^{66}$ as $\rho_q\eta^3\simeq 0.796$.  A series of histograms were obtained for each density by
carrying out simulations for different cylinder sizes   $L$ (and $R$, with $\omega=L/R=const=1.5$).  Equation
(25) was then applied  to each series and the surface tension $\sigma$ for an infinite system was found by 
extrapolating the series to  $\ln R/R^2 \rightarrow 0$ (see Figure 6): $\sigma\eta^2/kT=0.0037\pm 0.0005$ for
$\rho\eta^3=0.83$, and $\sigma\eta^2/kT=0.0063\pm 0.0008$ for $\rho\eta^3=0.85$.  
As expected, the surface tension increases with increasing density as the three phases become less and less
miscible.  Furthermore, according to the mean field approximation,$^{36}$  the line tension in this  model
above the quadruple point density is expected to be negative. Hence, prior to obtaining the final result, 
there is some uncertainty whether  condition (30) is fulfilled for a given $R$ (with $\omega=1.5$). The line
tension was thus determined by using both eq.(21) and eq.(33) and the results were substituted into eq.(30)
for verification. In neither case condition (30) held whence one can conclude that the dominant three-phase
partitioning is the one shown in Fig.4b, and eq.(33) is the appropriate one for determining the line tension.
The results for both densities (as shown in Figure 7) are: $\tau\eta/kT=-0.093 \pm 0.006$ for
$\rho\eta^3=0.83$ and  $\tau\eta/kT =-0.103\pm 0.008$ for $\rho\eta^3=0.85$. As expected,$^{36}$  the line
tension is  negative.  Relatively large standard deviations (about 10\% of average values) of results
for   $\sigma/kT$ and $\tau/kT$ are due to the uncertainty in finding the probabilities $p(o_1),\;p(o_2)$, and
$p(o_3)$ from a simulated histogram. To increase the accuracy of simulated $\sigma/kT$ and $\tau/kT$,
simulations longer than $2\times 10^7$ MC sweeps (a typical length of one simulation run in the present work) are
required.

\section{Conclusions}  A system where the three-phase equilibrium involves a three-phase contact region has
been studied. The main thermodynamic characteristic of a three-phase contact region  is the line tension,
defined as the linear  excess grand canonical potential per unit length of the contact line.  In such a
system, the probability distribution of states of the system with respect to two independent order parameters
depends on both the surface and line tensions. This constitutes the foundation of a method
for determining the line tension by MC simulations (either in a semi-grand canonical, or in a grand
canonical or in a Gibbs ensemble) proposed in the present work.   The method is the  combination of the
analysis of the probability distribution function  and finite-size scaling, and is the natural development of
Binder's approach$^{18}$ to  studying interfacial phenomena and phase equilibria in two-phase systems. 
MC 
simulations allow one to obtain the probability distribution of states of the system as a normalized
histogram. However, the histogram provided by MC simulations and results extracted therefrom  depend on the
size of the simulated system. Carrying out simulations for a series of system sizes, and extrapolating the
results obtained from the corresponding series of histograms to the  infinite size, one can determine the line
tension of the three-phase contact region and the  interfacial tensions of all three interfaces in an infinite
system (and hence the contact angles).  The proposed method is illustrated by its application to  the
three-dimensional  ternary fluid mixture, in which molecular pairs of like species do not interact whereas
those of unlike species interact as hard spheres. The three-phase equilibrium in this model above its
quadruple-point density involves a three-phase contact region. As expected, the interfacial tension increases
with increasing density. The behavior of the line tension is also compatible with what one would expect based
on the mean-field approximation,$^{36}$ although the predictions of the latter must be regarded with caution because
of the large discrepancy between the mean-field quadruple point density$^{47,48}$ ($\rho_q^{\mbox{\tiny
MF}}\eta^3\simeq 0.601$) and  the simulated one$^{66}$ ($\rho_q^{\mbox{\tiny MC}}\eta^3\simeq 0.796$). 

The present work suggests  that the most accurate data for the line tension can be obtained
by  simulations in a cylindrical cell under conditions where there is only one three-phase  contact line
in the cell. For a positive line tension this condition can be easily fulfilled by an appropriate choice
of the aspect ratio of the cylinder. For a negative line tension this condition may require the cylinder
to have not only a suitable aspect ratio but also a large enough radius which may lead to a drastic
increase of simulation time. In this case it may be a better choice to conduct simulations under
conditions where there are two three-phase contact lines in the simulation cell. However, in this case
the accuracy of results for the line tension can suffer if the contact lines are so close that the
inhomogeneities in two three-phase contact regions overlap. In order to avoid this effect it is necessary
to take longer cylinders which again leads to an increase in simulation time. 

It follows from the foregoing that it is possible to determine the line tension by simulations in a cubic cell
with PBC in two directions, say $x$ and $y$, and anti-PBC in the third, say $z$ (the extension of our method
to this case will be given in a sequel$^{67}$ to the present work). Moreover, combining such simulations with
the canonical ensemble MC simulations, carried out under  the same boundary conditions and appropriate choice
of the order parameters, makes it possible  to verify the validity of the new linear adsorption
equation$^{39}$  $d\tau=-\sum_i\Lambda_i\De\mu_i+c_{\alpha}d\alpha+c_{\beta}d\beta+c_{\gamma}d\gamma$, where 
$d\tau$ is    the change in the line tension due to the changes  $d \mu_i\;\;\;(i=1,2,3)$ of the field
variables  $\mu_i$ and $\Lambda_i$ is the linear adsorption of component $i$. The last three terms
($c_{\alpha}d\alpha$, etc) on the RHS of the above equation  make a difference between the old$^{2,19}$ and
modified$^{39}$  versions of the linear adsorption equation. At present, there are no recipes to analytically
evaluate the coefficients $c_{\alpha},c_{\beta},c_{\gamma}$. However,  the proposed method enables one to find
$d\tau$, while the subsequent canonical MC simulations can provide $\sum_i\Lambda_id\mu_i$. The inequality
$d\tau+\sum_i\Lambda_i\De\mu_i\ne 0$ could be regarded as an indirect ``simulational'' confirmation of the
validity of the modified linear adsorption equation (this will also be discussed in the sequel.$^{67}$).
Recently,$^{40}$ this inequality was shown to hold in the framework of  the square-gradient approximation of
density-functional theory with a special choice for the local part of the free-energy density.  Note that in
the model system, simulated in the present work, $d\alpha=d\beta=d\gamma=0$ for any changes of field
variables.  Thus, in this particular model  $d\tau=-\sum_i\Lambda_i\De\mu_i$, so  another model is needed to
simulate the validity of the new linear adsorption equation. 

Finally, it should be noted that the proposed method  can be easily modified to study wetting phenomena and,
in particular, to  determine the line tension attributed to the region of contact of two equilibrium  fluid
phases with an ``inert" rigid substrate.  As in the case of two-phase equilibrium discussed in section 2, the
probability distribution of states of such a system is a function of only one independent order parameter, but
the two maxima of this function (histogram)  will now depend on the corresponding fluid-substrate interfacial 
tensions and hence may not be equal, unlike the situation described in section 2 and Figure 1.  Furthermore,
the minimum of this function will now depend on the line tension associated with  the ``fluid-fluid-substrate"
contact region and may not be located  strictly in the middle between the two maxima. A
statistical-mechanical  basis for determining the line tension in such a system by MC simulations will be
given in a sequel$^{67}$ to the present work.

\subsubsection*{Acknowledgements}
The author is grateful to B. Widom for many helpful discussions and reading the manuscript of this work, 
which  was done as a part of Widom's  research program at Cornell University. It was supported by the U.S.
National Science Foundation and the Cornell Center for Material Research.

\appendix
\section*{Appendix.  Explicit form of auxiliary functions $s(\omega)$ and $\phi(\omega)$}
\newcounter{s2e}
\newcommand{\ale}{\setcounter{s2e}{\value{equation}}%
\stepcounter{s2e}\setcounter{equation}{0}\renewcommand{\theequation}{A\arabic{equation}}}
\ale
The function $s(\omega)$ in inequality (30) is defined as 
\be s(\omega)&=&-\frac1{3}(3+\pi)\omega+\frac6{\sqrt{3}}\pi-\frac4{\sqrt{3}}g(\omega)
-\frac{12}{\sqrt{3}}\arccos{g(\omega)}\nonumber\\
&&+\frac8{\sqrt{3}}g(\omega)\sqrt{a-g^2(\omega)}
-\frac3{2\sqrt{3}\pi g(\omega)}[\pi-2\arccos{g(\omega)}]^2 \;\;\;(1<\omega\le\omega'),\ee
\beq s(\omega)=-\frac1{3}(5+2\pi)+\frac{13}{2\sqrt{3}}\pi-\frac8{3\sqrt{3}}-
\frac{16}{3\sqrt{3}\pi} \;\;\;(\omega'<\omega<\pi),\eeq
where $g(\omega)$ is the solution of the equation 
\beq \frac{3}{4}\pi g(\omega)+\frac1{2}(1-g(\omega))Q(g(\omega))=\frac{\sqrt{3}}{4}\pi\omega+1\eeq
with the function $Q(x)$ of some variable $x$ being defined as 
\beq Q(x)=\frac{3\sin{\arccos{x}-}3x\arccos{x}-3\sin^3{\arccos{x}}}{1-x}.\eeq

\par The function $\phi(\omega)$ in eq.(32) is defined as 
\be \phi(\omega)&=&(2+\pi)\omega+\frac6{\sqrt{3}}\pi-\frac4{\sqrt{3}}g(\omega)
-\frac12{\sqrt{3}}\arccos{g(\omega)}\nonumber\\
&&+\frac8{\sqrt{3}}g(\omega)\sqrt{a-g^2(\omega)}
-\frac3{2\sqrt{3}\pi g(\omega)}[\pi-2\arccos{g(\omega)}]^2 \;\;\;(1<\omega\le\omega'),\ee
\beq \phi(\omega)=\frac2{3}(2+\pi)+\frac{13}{2\sqrt{3}}\pi-\frac8{3\sqrt{3}}-
\frac{16}{3\sqrt{3}\pi} \;\;\;(\omega'<\omega<\pi),\eeq
with $g(\omega)$ defined by eq.(A3). 
\section*{References}
\begin{list}{} 
\item $^{1}$ J. G. Kirkwood and F. P. Buff, J. Chem. Phys. 17, 338 (1949).
\item $^{2}$J.S. Rowlinson and B. Widom, {\it Molecular Theory of Capillarity} 
(Clarendon Press, Oxford, 1982).
\item $^{3}$ R. Evans, Adv.Phys. {\bf 28}, 143 (1979).
\item $^{3}$ R. Evans, {\it Fundamentals of inhomogeneous fluids}, edited by D. Henderson (Marcel Dekker,
NY, 1992), pp.85-175.
\item $^{5}$ B. Widom, {\it Phase Transitions and Critical Phenomena}, edited by C. Domb and M.S. Green
(Academic, New York, 1972), vol.II, pp.79-100.
\item $^{6}$ F.P. Buff, R.A. Lovett, and F.H. Stillinger,Jr., Phys. Ref. Lett. {\bf 15}, 621 (1965).
\item $^{7}$ J. K. Lee, J. A. Barker, and G. M. Pound, J. Chem. Phys. {\bf 60}, 1976 (1974).
\item $^{8}$ M. J. P. Nijmeijer, A. F. Bakker, C. Bruin, and J. H. Sikkenk, J. Chem. Phys. {\bf 89}, 3789 
(1988).
\item $^{9}$J. A. Barker, Mol. Phys. {\bf 80}, 815 (1993).
\item $^{10}$C. D. Holcomb, P. Clancy, and J. A. Zollweg, Mol. Phys. {\bf 78}, 437 (1993).
\item $^{11}$J. Miyazaki, J. A. Barker, and G. M. Pound, J. Chem. Phys.  {\bf 64}, 3364 (1976).
\item $^{12}$E. Salomons and M. Mareschal, J. Phys. Condensed Matter  {\bf 3}, 3645 (1991).
\item $^{13}$J. Potvin and C. Rebbi, Phys. Rev. Lett. {\bf 62}, 3062 (1989).
\item $^{14}$H. Gausterer, J. Potvin, C. Rebbi, and S. Sanielevici, Physica A {\bf 192}, 525 (1993).
\item $^{15}$J. E. Hunter III, W. P. Reinhardt, and T. F. Davis, J. Chem. Phys. {\bf 99}, 6856 (1993).
\item $^{16}$K. K. Mon and D. Jasnow, Phys. Rev. A {\bf 30}, 670 (1984);
{\it ibid.} Phys. Rev. A {\bf 31}, 4008 (1985). 
\item $^{17}$M. Hasenbusch, J. Phys. I France {\bf 3}, 753 (1993).
\item $^{18}$K. Binder, Phys. Rev. A {\bf 25}, 1699 (1982).
\item $^{19}$K. K. Mon and K. Binder, J. Chem. Phys. {\bf 96}, 6989 (1992).
\item $^{20}$B. Grossmann and M. L. Laursen, Nucl. Phys. B {\bf 408}, 637 (1993).
\item $^{21}$B. A. Berg, U. Hansmann, and T. Neuhaus, Phys. Rev. B {\bf 47}, 497 (1993).
\item $^{22}$B. A. Berg, U. Hansmann, and T. Neuhaus, Z. Phys. B {\bf 90}, 229 (1993).
\item $^{23}$B. A. Berg and T. Neuhaus, Phys. Rev. Lett. {\bf 68}, 9 (1992).
\item $^{24}$A. Billoire, T. Neuhaus, and B. A. Berg, Nucl. Phys. B {\bf 413}, 795 (1994).
\item $^{25}$W. Janke, B. A. Berg, and M. Katoot, Nucl. Phys. B {\bf 382}, 649 (1992).
\item $^{26}$N. A. Alves, Phys. Rev. D {\bf 46}, 3678 (1992). 
\item $^{27}$J .E. Hunter III and W.P. Reinhardt, J. Chem. Phys. {\bf 103}, 8627 (1995).
\item $^{28}$J. J. Potoff and A.Z. Panagiotopoulos, J. Chem. Phys. {\bf 112}, 6411 (2000).
\item $^{29}$J. K. Singh, D.A. Kofke, J.R. Errington, J. Chem. Phys. {\bf 119}, 3405 (2003).
\item $^{30}$P. Virnau, M. M\"uller, L.G. MacDowell, and K. Binder, J. Chem. Phys. {\bf 121}, 2169 (2004).
\item $^{31}$R. L. C. Vink, J. Horbach, and  K. Binder, (submitted to ???, 2004) 
\item $^{32}$D. Platikanov, M. Nedyalkov, and A. Scheludko, J. Colloid Interface Sci. {\bf 75}, 612 (1980).
\item $^{33}$D. Platikanov, M. Nedyalkov, and V. Nasteva, J. Colloid Interface Sci. {\bf 75}, 620 (1980).
\item $^{34}$P.-H. Puech, N. Borghi, E. Karatekin, and F. Brochard-Wyart, Phys.Rev.Lett. {\bf 90}, 128304 (2003).
\item $^{35}$L. Boruvka and A.W. Neumann,  J. Chem. Phys. {\bf 66}, 5464 (1977).
\item $^{36}$J. Kerins and B. Widom, J. Chem. Phys. {\bf 77}, 2061 (1982).
\item $^{37}$I. Szleifer and B. Widom, Molec. Phys., {\bf 75}, 925 (1992).
\item $^{38}$B. Widom, Colloids Surf. A {\bf 239}, 141 (2004).
\item $^{39}$Y. Djikaev and B. Widom, J. Chem. Phys. {\bf 121}, 5602 (2004).
\item $^{40}$C. M. Taylor and B. Widom, (accepted for publication in JCP, 2004)
\item $^{41}$P. Tarazona and G. Navascu\'es, J. Chem. Phys. {\bf 75}, 3114 (1981).
\item $^{42}$M.J.P. Nijmeijer, C. Bruin, A.F. Baker, and J.M.J. Van Leeuwen, Phys. Rev. A {\bf 42}, 6052
(1990).
\item $^{43}$F. Bresme and N. Quirke, J. Chem. Phys.  {\bf 112}, 5985 (2000); 
{\it ibid}. J. Chem. Phys. {\bf 110}, 3536 (1999);
{\it ibid}. Phys. Chem. Chem. Phys.{\bf 1}, 2149 (1999);
{\it ibid}. Phys. Rev. Lett. {\bf 80}, 3791 (1998).
\item $^{44}$A. Patrykiejew, L. Salamacha, S. Sokoowski, and O. Pizio, Phys. Rev. E {\bf 67}, 061603
(2003).
\item $^{45}$A. Milchev, M. M\"uller, K. Binder, and D. P. Landau, Phys. Rev. E {\bf 68}, 031601 (2003).
\item $^{46}$D.R. Heine, G.S. Grest, and E. B. Webb III, Phys. Rev. E {\bf 70}, 011606 (2004).
\item $^{47}$M.I. Guerrero, J.S. Rowlinson, and G. Morrison, J. Chem. Soc. Faraday Trans. II {\bf 72}, 1970
(1976). 
\item $^{48}$N. Derosier, M.I. Guerrero, J.S. Rowlinson, and D. Stubley, 
J. Chem. Soc. Faraday Trans. II {\bf 73}, 1632 (1977). 
\item $^{49}$J.S.Rowlinson, Adv. Chem. Phys. {\bf 41}, 1 (1980).
\item $^{50}$M. Fitzgerald, R.R. Picard, and R.N. Silver, Europhys. Lett. {\bf 46}, 282 (1999).
\item $^{51}$T. Hill, {\it Statistical Mechanics} (McGraw-Hill, New York, 1956).
\item $^{52}$H.-P. Deutsch and K. Binder, Macromolecules {\bf 25}, 6214 (1992).
\item $^{53}$H.-P. Deutsch, J. Stat. Phys. {\bf 67}, 1039 (1992).
\item $^{54}$A. M. Ferrenberg and R. H. Swendsen, Phys. Rev. Lett. {\bf 61}, 2635 (1988).
\item $^{55}$K. Binder, Z. Phys. B {\bf 43}, 119 (1981).
\item $^{56}$B. Widom and J.S. Rowlinson,  J. Chem. Phys. {\bf 52}, 1670 (1970).
\item $^{57}$D. A. Kofke and E. D. Glandt, Mol. Phys. {\bf 64}, 1105 (1988).
\item $^{58}$D. Frenkel, {\it Computer Simulation in Chemical Phsysics}, edited by M.P. ALlen and D.J.
Tildesly (Kluwer Academic, New York, 1993), pp.93-152.
\item $^{59}$E. de Miguel, E.M. del Rio, and M.M. Telo da Gama, J. Chem. Phys. {\bf 103}, 6188 (1995).
\item $^{60}$C.-Y. Shew and A. Yethiraj, J. Chem. Phys. {\bf 104}, 7665 (1996).
\item $^{61}$K. Jagannathan and A. Yethiraj, J. Chem. Phys. {\bf 118}, 7907 (2003).
\item $^{62}$K. Binder, Phys. Rev. Lett. {\bf 47}, 693 (1982).
\item $^{63}$A. Z. Panagiotopoulos, Mol. Phys. {\bf 61}, 813 (1987).
\item $^{64}$A. Z. Panagiotopoulos, Mol. Simul. {\bf 9}, 1 (1992).
\item $^{65}$B. Smit, {\it Computer Simulation in Chemical Phsysics}, edited by M.P. ALlen and D.J.
Tildesly (Kluwer Academic, New York, 1993), pp.173-205. 
\item $^{66}$G. Johnson, H.Gould, J. Machta, and L.K. Chayes, Phys. Rev. Lett. {\bf 79}, 2612 (1997).
\item $^{67}$Y. Djikaev {\em Manuscript in preparation...}
\end{list}
\newpage
\section*{Captions to Figures 1-7}
of the manuscript {\bf } 
{\sc ``Histogram analysis as a method for determining the line tension 
by Monte-Carlo simulataions''} by Yuri Djikaev.
\subsubsection*{} 
{\bf Figure 1}: Typical ``histogram", probability distribution $p$ of states of a system with respect to a
single  order parameter $m$, under conditions of two-phase coexistence. The values of the order parameter
in two equilibrium phases are $m^{\alpha}$ and $m^{\beta}$ \vspace{0.17cm}\\ 
{\bf Figure 2}: Typical probability distribution $p$ of states of a system with respect to two 
order parameters $m_1$ and $m_2$ under conditions of three-phase coexistence, obtained for 
a three-dimensional 
ternary fluid mixture, in which molecular pairs of like species do not interact whereas those of unlike
species interact as hard spheres, at the total density $\rho\eta^3=0.85$. a) Three-dimensional
plot; b) Projection of the histogram onto the mole-fraction plane (for details see the text). \vspace{0.17cm}\\ 
{\bf Figure 3}:  Three two-phase (half-and-half) partitionings compatible with the periodic boundary
conditions along the $z$ axis: a) AL2P partitioning; b) droplet-like partitioning; c) transversal
partitioning. \vspace{0.17cm}\\ 
{\bf Figure 4}: a) The axial longitudinal three-phase (AL3P) partitioning in the cylindrical simulation cell
with the contact angles $\alpha,\;\beta$, and $\gamma$; b) and c) - the  
alternative to AL3P three-phase partitionings in the cilinder for $\omega>\omega'$ and 
$\omega<\omega'$ respectively  \vspace{0.17cm}\\ 
{\bf Figure 5}: Determination of the critical density for the primitive version of 
the two-dimensional Widom-Rowlinson model by SGCE MC simulations. 
The reduced second moment $\langle s^2\rangle/\langle |s|\rangle^2$ ($s=2m-1$ with $m$ the mole fraction of one of two components) 
is plotted vs dimensionless density $\rho\eta^2$ 
for four different numbers of molecules in the simulated system (square with PBC):
larger circles are for $N=256$, smaller points for $N=726$, smaller circles for $N=1024$, and larger points for 
$N=4098$. The thin curves are plotted only for guiding the eye. The dashed vertical line indicates 
the critical density from ref.66. \vspace{0.17cm}\\ 
{\bf Figure 6}: Determination of the surface tension by the histogram analysis combined with the 
finite size scaling. The LHS of eq.(25) is plotted vs $\ln{L}/L^2$ for two different densities,
$\rho\eta^3=0.83$ (upper series) and $\rho\eta^3=0.85$ (lower series). The straight lines show the linear 
least-squares fits to each series of simulated points. The extrapolated values for the infinite system size 
are: $\sigma\eta^2/kT=0.0037\pm 0.0005$ for
$\rho\eta^3=0.83$, and $\sigma\eta^2/kT=0.0063\pm 0.0008$ for $\rho\eta^3=0.85$.
\vspace{0.17cm}\\ 
{\bf Figure 7}: Determination of the surface tension by the histogram analysis combined with the 
finite size scaling. The LHS of eq.(33) is plotted vs $\ln{L}/L$ for two different densities,
$\rho\eta^3=0.83$ (upper series) and $\rho\eta^3=0.85$ (lower series). The straight lines show the linear 
least-squares fits to each series of simulated points. The extrapolated values for the infinite system size 
are: $\tau\eta/kT=-0.093\pm 0.006$ for $rho\eta^3=0.83 $, and $\tau\eta/kT=-0.103\pm 0.008$ for 
$rho\eta^3=0.83$. 
\vspace{0.17cm}\\ 
\begin{figure}[htp]
	\begin{flushleft}
              \leavevmode 
      	      \epsfxsize=27cm 
              \epsfbox{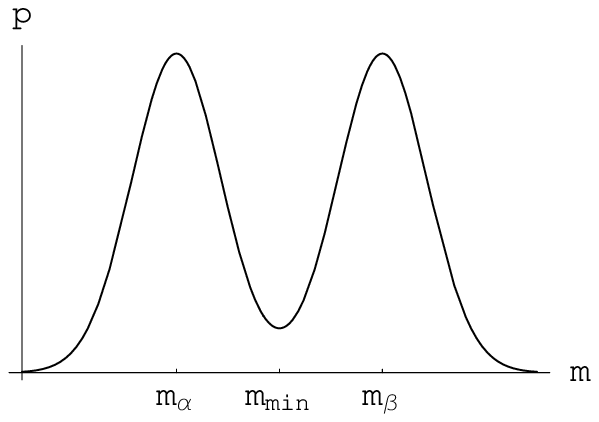}\\ \vspace{-15cm} 
		\caption{\small }
	\end{flushleft}
\end{figure}
\begin{figure}[htp]\vspace{-1cm}
	      \begin{center}
	      $\begin{array}{c@{\hspace{0.3cm}}c}
              \leavevmode
	\leavevmode\hbox{a)\vspace{1cm}} &  
      	      \epsfxsize=7.3cm
      	      \epsfbox{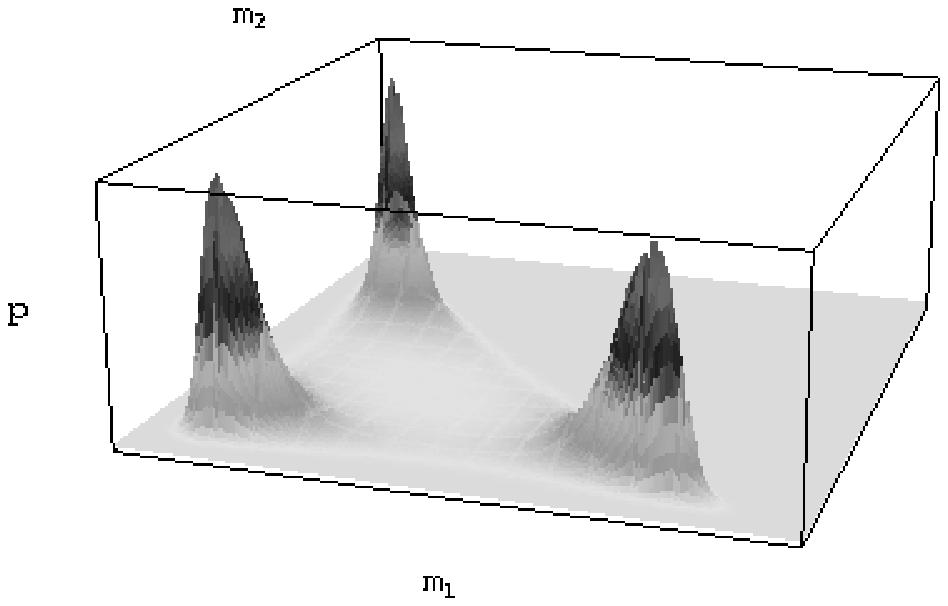} \\ [0.5cm]
	\leavevmode\hbox{b)} &  
      	      \epsfxsize=7.3cm
      	      \epsfbox{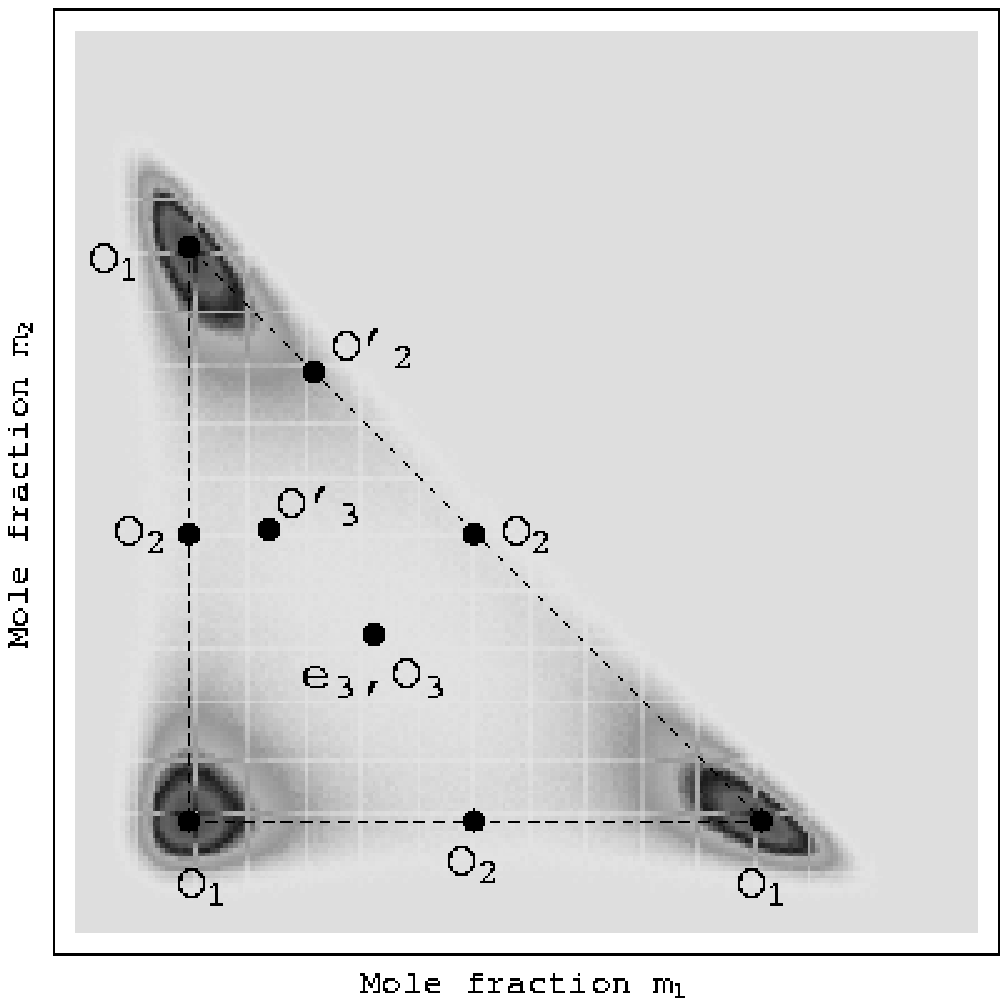} \\ [0.7cm]
\end{array}$
	      \end{center}
            \caption{\small }
\end{figure}\\
\begin{figure}[htp]\vspace{-1cm}
	      \begin{center}
	      $\begin{array}{c@{\hspace{1.5cm}}c}
              \leavevmode
	\leavevmode\hbox{a)} &  
      	      \epsfysize=5.9cm
      	      \epsfbox{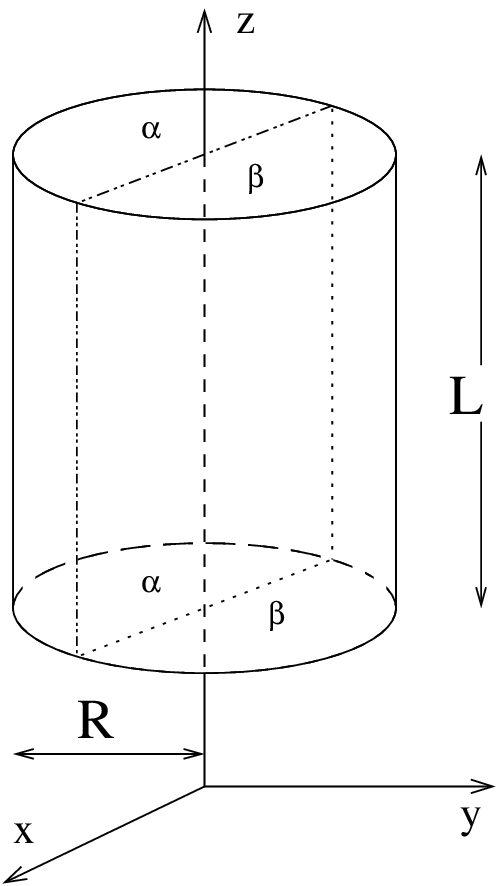} \\ [0.7cm]
	\leavevmode\hbox{b)} &  
      	      \epsfysize=5.9cm
      	      \epsfbox{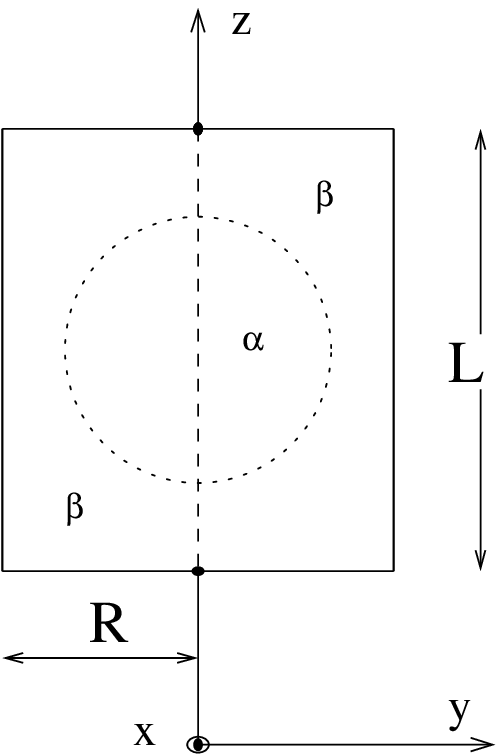} \\ [0.7cm]
 	\leavevmode\hbox{c)} &  
     	      \epsfysize=5.9cm
      	      \epsfbox{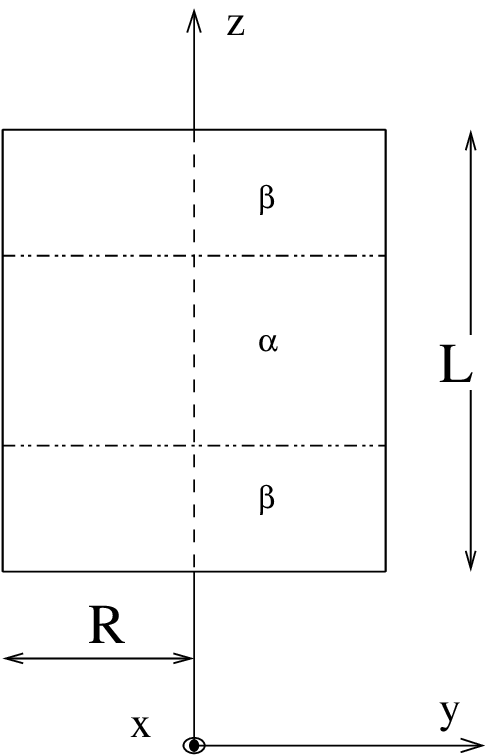}\\ [0.7cm]
\end{array}$
	      \end{center}
            \caption{\small }
\end{figure}\\
\begin{figure}[htp]\vspace{-1cm}
	      \begin{center}
	      $\begin{array}{c@{\hspace{1.5cm}}c}
              \leavevmode
	\leavevmode\hbox{a)} &  
      	      \epsfysize=5.9cm
      	      \epsfbox{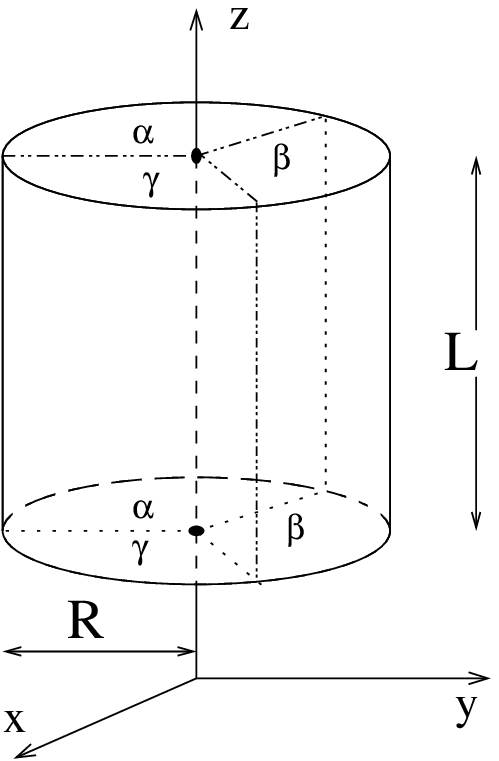} \\ [0.7cm]
	\leavevmode\hbox{b)} &  
      	      \epsfysize=5.9cm
      	      \epsfbox{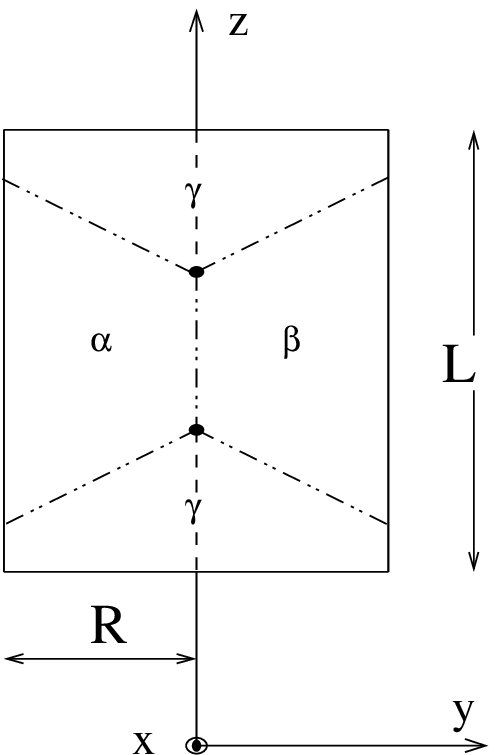} \\ [0.7cm]
 	\leavevmode\hbox{c)} &  
     	      \epsfysize=5.9cm
      	      \epsfbox{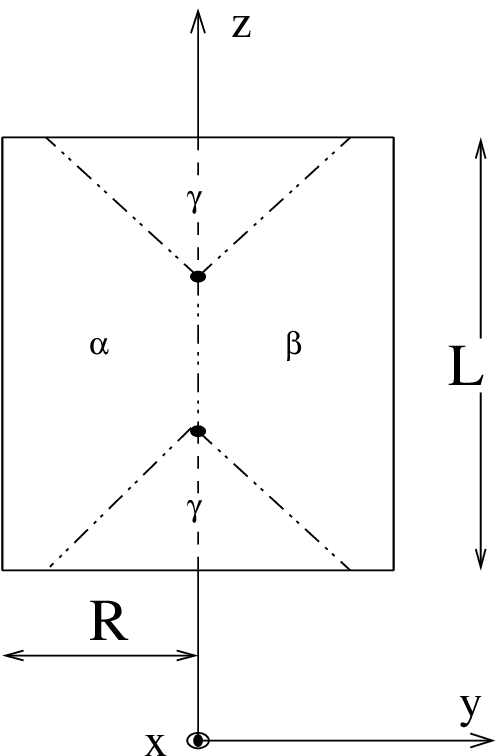}\\ [0.7cm]
\end{array}$
	      \end{center}
            \caption{\small }
\end{figure}\\
\begin{figure}[htp]
              \begin{center}
              \leavevmode
      	      \epsfxsize=21.9cm\vspace{-2cm}
              \epsfbox{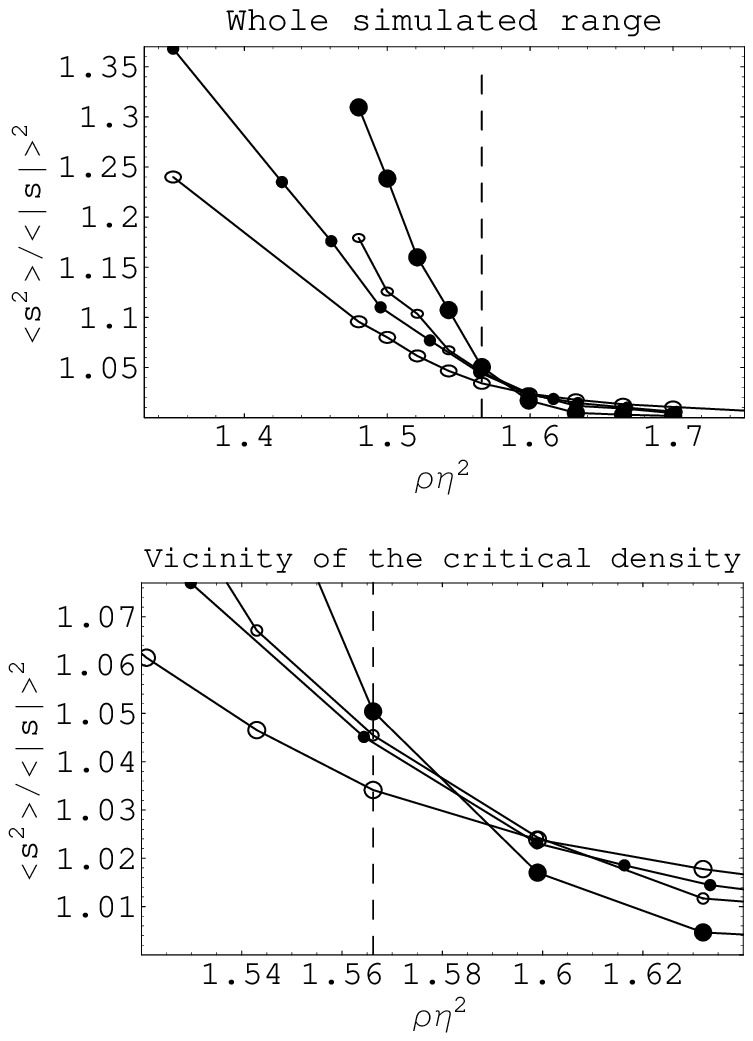}\vspace{-10cm}
              \end{center}
              \caption{}
\end{figure}
\begin{figure}[htp]
              \begin{center}
              \leavevmode
      	      \epsfxsize=22.3cm
              \epsfbox{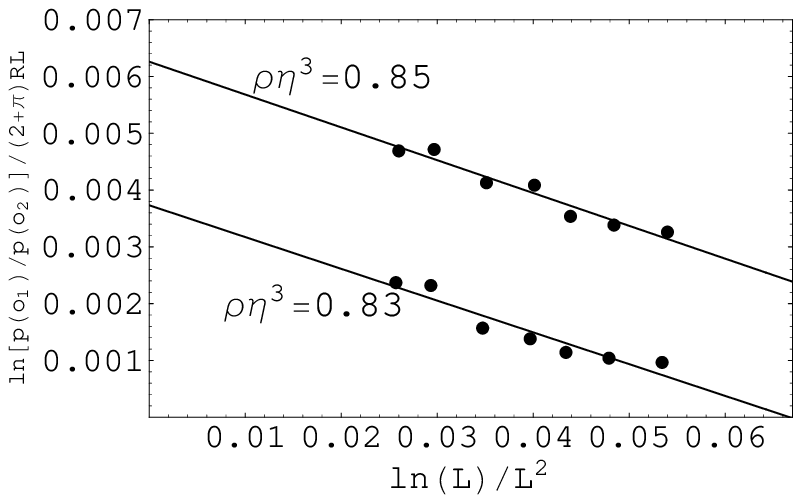}\vspace{-10cm}
              \end{center}
              \caption{ }
\end{figure}
\newpage
\begin{figure}[tp]
              \begin{center}
              \leavevmode
      	      \epsfxsize=21.9cm
              \epsfbox{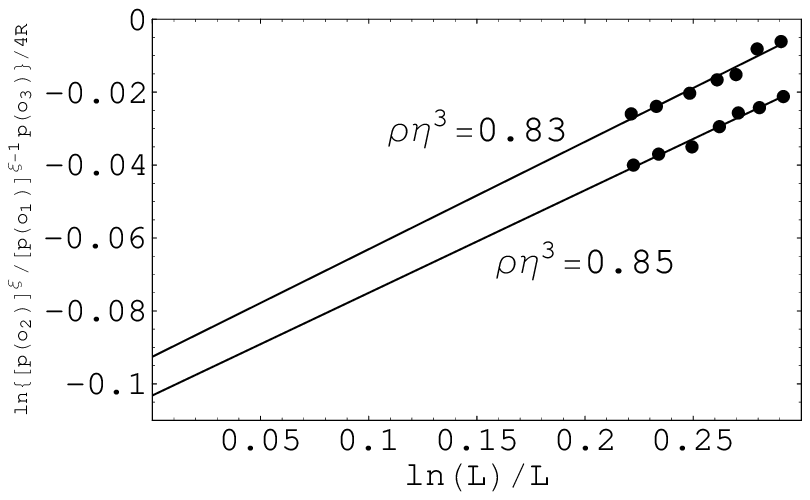}\vspace{-10cm}
              \caption{\large }
              \end{center}
\end{figure}
\end{document}